\documentclass[twocolumn,aps,pra]{revtex4-2} 
\usepackage{bm}    
\usepackage{amssymb,amsmath}
\usepackage{times}
\usepackage{graphicx}

\begin{document}

\title{Physical relevance of time-independent scattering calculations in non-Hermitian systems: \\The role of time-growing bound states}

\author{Chao Zheng}
\email{zhengchaonju@gmail.com}
\affiliation{School of Education, Jiangsu Open University, Nanjing 210036, China}

\begin{abstract}
Time-independent scattering methods are widely employed to analyze transport in non-Hermitian systems. 
Their application, however, rests on a critical yet often overlooked assumption: that an incident wave is a pure superposition of scattering states. 
In practice, any physically realistic, spatially localized wave packet will generally have a nonzero overlap with the system's bound states, thereby violating this premise.
While this violation is inconsequential in Hermitian systems, it can invalidate the conventional scattering picture in their non-Hermitian counterparts.
The underlying cause is the emergence of time-growing bound states, which manifest as poles of the scattering matrix ($S$ matrix) in the first quadrant of the complex wave-number plane.
Any initial overlap with these states becomes exponentially amplified, eventually dominating the long-time dynamics.
Consequently, the actual evolution of a wave packet diverges dramatically from the conventional scattering picture, rendering the transmission and reflection coefficients derived from time-independent scattering methods unphysical.
Using tight-binding models with non-Hermiticity introduced via imaginary on-site potentials or asymmetric hopping, we demonstrate that parameter regimes supporting such growing states are common. 
We therefore conclude that an analysis of $S$-matrix poles is an indispensable step to confirm the physical relevance of time-independent scattering calculations in non-Hermitian systems.
\end{abstract}

\date{\today}

\maketitle

\section{Introduction}
\label{sec:Introduction}

Non-Hermitian systems have attracted significant attention in recent years, offering a unique platform for exploring fundamental physical phenomena and developing novel technological applications~\cite{el-ganainy2018,ozdemir2019,ashida2020,bergholtz2021}. 
Unlike their Hermitian counterparts, which describe closed systems with real energy eigenvalues, non-Hermitian systems characterize open systems that exchange energy or particles with their environment, resulting in complex energy spectra~\cite{rotter2009,moiseyev2011}.
Such non-Hermiticity gives rise to a plethora of intriguing phenomena absent in Hermitian systems, including exceptional points~\cite{bender2007,heiss2012,ozdemir2019,miri2019}, the non-Hermitian skin effect~\cite{lee2016,yao2018,kunst2018,zhang2022a}, and exotic topological phases~\cite{shen2018,yao2018a,takata2018,gong2018,kawabata2019,liu2019a}. 
The rapid development of this field is driven by both theoretical breakthroughs and experimental advances across diverse areas, including photonics~\cite{feng2017,ozdemir2019}, acoustics~\cite{zhu2014,fleury2015,cummer2016,auregan2017}, and condensed-matter physics~\cite{chen2021,chen2022,abbasi2022,li2023}.

One of the key aspects of studying non-Hermitian systems is understanding their scattering properties~\cite{muga2004,jones2007,dogan2008,znojil2008,ahmed2012,li2015,ahmed2018,ruschhaupt2018,burke2020,jin2021}.
Scattering describes how waves interact with a system and are subsequently reflected or transmitted.
Key quantities of interest include reflection and transmission amplitudes, as well as the scattering matrix ($S$ matrix), which encapsulates the full scattering information.
$S$-matrix poles in the complex energy or wave-number $k$ plane are particularly important, as they reveal system features such as resonances and bound states~\cite{moiseyev2011}.
In Hermitian systems, scattering is typically described by a unitary $S$ matrix, ensuring probability conservation. 
However, in non-Hermitian systems, the $S$ matrix is generally nonunitary, reflecting the possibility of amplification or attenuation of scattered waves. 
This nonunitarity gives rise to unique scattering phenomena, including unidirectional invisibility~\cite{lin2011,ge2012,feng2013,fleury2015}, spectral singularities~\cite{mostafazadeh2009,longhi2009,ramezani2014,wang2016,jin2018b}, coherent perfect absorption~\cite{longhi2010,chong2010,wan2011,baranov2017}, and robust unidirectional transport~\cite{longhi2015,longhi2015a}.
These intriguing scattering properties pave the way for unprecedented control over wave propagation and novel devices with functionalities unattainable in Hermitian systems.

Although the scattering process is inherently time-dependent, the transport properties of both Hermitian and non-Hermitian systems are often analyzed using time-independent scattering methods~\cite{datta1997,beenakker1997,markos2008,schomerus2013}, such as the transfer-matrix method~\cite{mostafazadeh2009,lin2011,ramezani2014,auregan2017,achilleos2017a,luo2021}.
Their application, however, rests on a critical yet often overlooked assumption: that an incident wave is a pure superposition of scattering states~\cite{taylor1972}. 
In practice, any initial localized wave packet, even when prepared far from the scattering region, will generally have a nonzero overlap with the evanescent tails of the system's bound states, thereby violating this premise.
While this violation is inconsequential in Hermitian systems, where bound states are temporally stable, we will demonstrate that its violation in non-Hermitian systems can lead to a breakdown of the conventional scattering picture.

The potential consequences of this violation were first noted in random laser studies. 
Researchers observed that for waves propagating through gain layers, time-independent calculations could yield unphysical transmission coefficients when the gain or system size exceeded certain thresholds~\cite{jiang1999,ma2001,bahlouli2005}. 
Despite these early warnings, the issue has gained renewed urgency with the recent surge of interest in modern non-Hermitian systems.
Prominent examples include parity-time ($\mathcal{PT}$) symmetric systems~\cite{vazquez-candanedo2014,garmon2015,zhu2016,achilleos2017a,shobe2021}, anti-$\mathcal{PT}$-symmetric systems~\cite{ge2013,peng2016,choi2018,jin2018a,xu2021,xu2023a,wu2021a,jangjan2024}, asymmetric hopping systems~\cite{zhang2013a,longhi2015a,longhi2015,li2021}, and various non-Hermitian disordered systems~\cite{basiri2014,luo2021,liu2021a,liu2021,schiffer2021,tang2021,tzortzakakis2021,kawabata2021,huang2022,molignini2023,li2024,halder2025}.
Many of these systems are formulated using tight-binding models, where non-Hermiticity is introduced through two distinct mechanisms: imaginary on-site potentials and asymmetric hopping terms~\cite{hatano1996,hatano1997}.  
The former are discrete analogs of the continuous gain-and-loss layers where the problem was first identified. 
In contrast, the latter mechanism, asymmetric hopping, has no direct counterpart in continuous systems, which makes the applicability of time-independent scattering calculations less obvious.
Nevertheless, most studies on these modern non-Hermitian systems have employed time-independent scattering methods to analyze their transport properties, implicitly assuming the validity of the conventional scattering picture.
This practice risks producing theoretical predictions that are physically irrelevant to the actual evolution of a wave packet, potentially leading to erroneous conclusions about the underlying physics~\cite{vazquez-candanedo2014,zhu2016,shobe2021,xu2021,li2021,luo2021}.

In this work, we systematically investigate the conditions under which time-independent scattering calculations yield a physically relevant description of wave-packet dynamics in non-Hermitian systems.
We focus on tight-binding models with non-Hermiticity introduced through imaginary on-site potentials or asymmetric hopping terms.
By violating the fundamental assumption of the time-independent scattering method, the actual wave-packet evolution can diverge drastically from the method's predictions.
Through direct comparison with time-dependent wave-packet simulations, we identify the parameter regimes where this divergence occurs.
This divergence originates from the emergence of time-growing bound states, which manifest as $S$-matrix poles in the first quadrant of the complex $k$ plane~\cite{bahlouli2005}.
Although the wave packet's initial overlap with these states may be negligible, this component undergoes exponential amplification, eventually dominating the system's long-time dynamics. 
Consequently, the wave-packet evolution deviates dramatically from the conventional scattering picture, rendering the transmission and reflection coefficients obtained from time-independent scattering methods unphysical.
We demonstrate that this phenomenon is a general feature across various non-Hermitian tight-binding models.
We therefore conclude that an analysis of $S$-matrix poles is indispensable for confirming the physical relevance of time-independent scattering calculations in non-Hermitian systems.

The remainder of this paper is organized as follows. 
In Sec.~\ref{sec:Imaginary_on_site_potential_Model}, we consider a model with imaginary on-site potentials. 
We first analyze this model using the time-independent scattering method (Sec.~\ref{sec:Time_independent_method}) and then compare the results with full time-dependent wave-packet simulations (Sec.~\ref{sec:Time_dependent_method}). 
The emergence of time-growing bound states is explored first through an eigenvalue analysis of the finite system (Sec.~\ref{sec:Eigenvalues_and_eigenstates_of_the_finite_system}) and then more rigorously through an $S$-matrix pole analysis of the infinite system (Sec.~\ref{sec:S_matrix_poles}).
Section~\ref{sec:Physical_relevance_of_time_independent_scattering_calculations_in_non_Hermitian_systems} consolidates these findings to explain the physical relevance of time-independent scattering calculations in non-Hermitian systems.
In Sec.~\ref{sec:Asymmetric_hopping_models}, we extend our investigation to non-Hermitian systems with asymmetric hopping terms, examining four distinct models: an unequal hopping model (Sec.~\ref{sec:Unequal_hopping_model}), a complex hopping model (Sec.~\ref{sec:Complex_hopping_model}), an anti-Hermitian hopping model (Sec.~\ref{sec:Anti_Hermitian_hopping_model}), and an imaginary coupling model (Sec.~\ref{sec:Imaginary_coupling_model}). 
Finally, we summarize and discuss our findings in Sec.~\ref{sec:Discussion}.

\section{Imaginary on-site potential Model}
\label{sec:Imaginary_on_site_potential_Model}

\begin{figure}
\centering
\includegraphics[width=0.95\columnwidth]{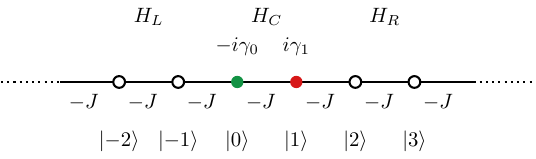}
\caption{
Schematic illustration of a non-Hermitian scattering system with gain and loss.
The system consists of a scattering center coupled to two semi-infinite tight-binding chains.
}
\label{fig:00_Schematic}
\end{figure}

We begin by considering a one-dimensional (1D) tight-binding model with imaginary on-site potentials, as illustrated in Fig.~\ref{fig:00_Schematic}.
The system consists of a scattering center coupled to two semi-infinite leads. 
The Hamiltonian of the system is given by
\begin{equation}
H=H_{L} +H_{C} +H_{R},
\label{eq:full_Hamiltonian}
\end{equation}
where
\begin{equation}
H_{L} =-J\sum _{j=-\infty }^{-1} (|j\rangle \langle j+1|+\mathrm{H.c.} ),
\end{equation}
\begin{equation}
H_{R} =-J\sum _{j=2}^{\infty } (|j\rangle \langle j-1|+\mathrm{H.c.} ),
\end{equation}
\begin{equation}
H_{C} =-J(|0\rangle \langle 1|+|1\rangle \langle 0|)-i\gamma _{0} |0\rangle \langle 0|+i\gamma _{1} |1\rangle \langle 1|
\end{equation}
describe the left lead, right lead, and scattering center, respectively.
Here, $|j\rangle$ represents the Wannier state localized at site $j$,
$J$ denotes the hopping strength between adjacent sites, and $\gamma_0$ and $\gamma_1$ are positive parameters characterizing the gain and loss strengths at sites 0 and 1, respectively.
The Hamiltonian exhibits $\mathcal{PT}$ symmetry when $\gamma_0 = \gamma_1$. 
In this study, we consider general cases where $\gamma_0$ and $\gamma_1$ can assume arbitrary values.
For simplicity, we set $\hbar$, the hopping strength $J$, and the lattice constant $a$ to unity, thereby expressing energies and times in units of $J$ and $\hbar /J$, and lengths and wave numbers in units of $a$ and $1/a$, respectively.

\subsection{Time-independent scattering method}
\label{sec:Time_independent_method}

We first analyze the scattering properties of this system using the time-independent scattering method.
This approach determines the reflection and transmission coefficients from the stationary scattering states, which are the real-energy eigenstates of the Hamiltonian.
These states are obtained by solving the time-independent Schrödinger equation
\begin{equation}
H|\psi\rangle = E |\psi\rangle,
\end{equation}
subject to specific scattering boundary conditions.
Expanding the wave function in the Wannier basis, $|\psi \rangle =\sum _{j} \psi (j)|j\rangle$,
transforms the Schrödinger equation into coupled equations for the coefficients $\psi (j)$.
For the leads ($j \leq -1$ and $j \geq 2$), the equation takes the form
\begin{equation}
-\psi (j-1)-\psi (j+1)=E\psi (j).
\label{eq:Schrodinger_equation_leads}
\end{equation}
For the scattering center ($j=0$ and 1), we have
\begin{align}
-\psi (-1)-i\gamma _{0} \psi (0)-\psi (1) & =E\psi (0),\label{eq:Schrodinger_equation_scattering_center_site0}\\
-\psi (0)+i\gamma _{1} \psi (1)-\psi (2) & =E\psi (1).\label{eq:Schrodinger_equation_scattering_center_site1}
\end{align}
In the leads, solutions to Eq.~(\ref{eq:Schrodinger_equation_leads}) 
 are linear combinations of plane waves $e^{ikj}$ and $e^{-ikj}$, where the wave number $k$ and energy $E$ are related by the dispersion relation
\begin{equation}
E = -2\cos k.
\end{equation}
This relation defines the scattering continuum of the system, which spans the energy range $E \in [-2, 2]$.

The scattering properties are encoded in two linearly independent solutions, corresponding to waves incident from the left and right.
The first eigenstate $\psi_{L}^{k} (j)$ represents a wave incident from the left
\begin{equation}
\psi _{L}^{k} (j)=\begin{cases}
A e^{ikj} +B e^{-ikj} & \text{for } j \leq -1,\\
\psi _{L}^{k} (0) & \text{for } j=0,\\
\psi _{L}^{k} (1) & \text{for } j=1,\\
C e^{ikj} & \text{for } j \geq 2,
\end{cases}
\label{eq:wave_function_left_scattering}
\end{equation}
where $0 \leq k \leq \pi$. 
The reflection and transmission amplitudes for left incidence are given by $r_L = B/A$ and $t_L =C/A$, respectively.
The second eigenstate $\psi _{R}^{k} (j)$ represents a wave incident from the right
\begin{equation}
\psi _{R}^{k} (j)=\begin{cases}
B e^{-ikj} & \text{for } j \leq -1,\\
\psi _{R}^{k} (0) & \text{for } j=0,\\
\psi _{R}^{k} (1) & \text{for } j=1,\\
C e^{ikj} + D e^{-ikj} & \text{for } j \geq 2,
\end{cases}
\label{eq:wave_function_right_scattering}
\end{equation}
where $0 \leq k \leq \pi$. 
The reflection and transmission amplitudes for right incidence are $r_R=C/D$ and $t_R=B/D$, respectively.

For left-incident scattering, the Schrödinger equations at the lead sites adjacent to the scattering center ($j=-1$ and 2) are given by
\begin{equation}
\begin{aligned}
-\psi _{L}^{k} (-2)-\psi _{L}^{k} (0) & =E\psi _{L}^{k} (-1),\\
-\psi _{L}^{k} (1)-\psi _{L}^{k} (3) & =E\psi _{L}^{k} (2).
\end{aligned}
\end{equation}
Substituting the wave function expressions from Eq.~(\ref{eq:wave_function_left_scattering}) and solving for $\psi_L^k(0)$ and $\psi_L^k(1)$ gives
\begin{equation}
\psi _{L}^{k} (0)=A+B ,\quad \psi _{L}^{k} (1)=C e^{ik} .
\end{equation}
Substituting these expressions into the Schrödinger equations for the scattering center [Eqs.~(\ref{eq:Schrodinger_equation_scattering_center_site0}) and (\ref{eq:Schrodinger_equation_scattering_center_site1})] enables us to solve for $B$ and $C$:
\begin{align}
B & =A\frac{-\gamma _{0} +\gamma _{1} e^{2ik} -i\gamma _{0} \gamma _{1} e^{ik}}{\gamma _{0} -\gamma _{1} +i\gamma _{0} \gamma _{1} e^{ik} +2\sin k},\\
C & =A\frac{2\sin k}{\gamma _{0} -\gamma _{1} +i\gamma _{0} \gamma _{1} e^{ik} +2\sin k}.
\end{align}
Consequently, the reflection and transmission amplitudes for left incidence are
\begin{align}
r_{L} & =\frac{B}{A} =\frac{-\gamma _{0} +\gamma _{1} e^{2ik} -i\gamma _{0} \gamma _{1} e^{ik}}{\gamma _{0} -\gamma _{1} +i\gamma _{0} \gamma _{1} e^{ik} +2\sin k}, \label{eq:rL}\\
t_{L} & =\frac{C}{A} =\frac{2\sin k}{\gamma _{0} -\gamma _{1} +i\gamma _{0} \gamma _{1} e^{ik} +2\sin k}. \label{eq:tL}
\end{align}

Similarly, for right-incident scattering, the Schrödinger equations for sites $j=-1$ and $2$ take the form
\begin{equation}
\begin{aligned}
-\psi _{R}^{k} (-2)-\psi _{R}^{k} (0) & =E\psi _{R}^{k} (-1),\\
-\psi _{R}^{k} (1)-\psi _{R}^{k} (3) & =E\psi _{R}^{k} (2).
\end{aligned}
\end{equation}
Substituting the wave function expressions from Eq.~(\ref{eq:wave_function_right_scattering}), we find
\begin{equation}
\psi _{R}^{k} (0)=B ,\quad \psi _{R}^{k} (1)= C e^{ik} + D e^{-ik}.
\end{equation}
Substituting these into Eqs.~(\ref{eq:Schrodinger_equation_scattering_center_site0}) and (\ref{eq:Schrodinger_equation_scattering_center_site1}), and solving for $B$ and $C$ gives
\begin{align}
C & =D\frac{-\gamma _{0} +\gamma _{1} e^{-2ik} -i\gamma _{0} \gamma _{1} e^{-ik}}{\gamma _{0} -\gamma _{1} +i\gamma _{0} \gamma _{1} e^{ik} +2\sin k} ,\\
B & =D\frac{2\sin k}{\gamma _{0} -\gamma _{1} +i\gamma _{0} \gamma _{1} e^{ik} +2\sin k} .
\end{align}
The reflection and transmission amplitudes for right incidence are thus
\begin{align}
r_{R} & =\frac{C}{D} =\frac{-\gamma _{0} +\gamma _{1} e^{-2ik} -i\gamma _{0} \gamma _{1} e^{-ik}}{\gamma _{0} -\gamma _{1} +i\gamma _{0} \gamma _{1} e^{ik} +2\sin k} ,\label{eq:rR}\\
t_{R} & =\frac{B}{D} =\frac{2\sin k}{\gamma _{0} -\gamma _{1} +i\gamma _{0} \gamma _{1} e^{ik} +2\sin k} .\label{eq:tR}
\end{align}
The reflection and transmission coefficients are given by $R_{L(R)} :=|r_{L(R)} |^{2}$ and $T_{L(R)} :=|t_{L(R)} |^{2}$, respectively.

\begin{figure}
\centering
\includegraphics[width=\columnwidth]{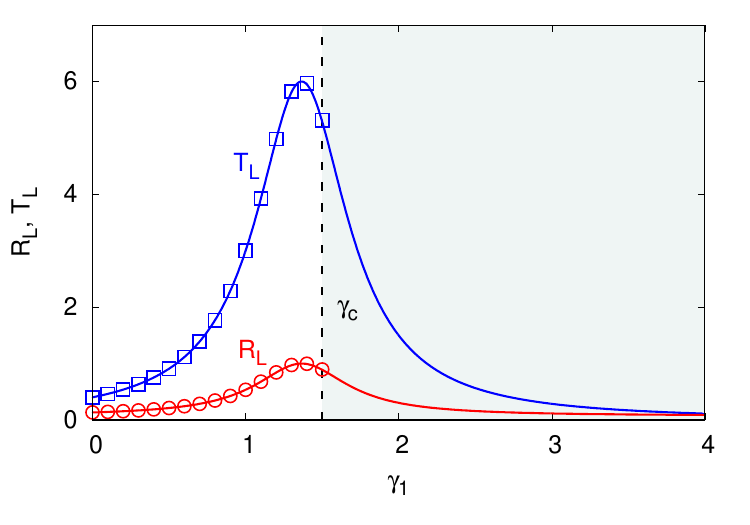}
\caption{
Reflection and transmission coefficients ($R_L$ and $T_L$) for left-incident waves as a function of $\gamma_1$, with fixed parameters $\gamma_0 = 1$ and $k = \pi/3$.
Solid lines represent results from the time-independent scattering method, while circles and squares depict results from wave-packet evolution simulations.
A significant divergence between the two approaches is observed when $\gamma_1$ exceeds the critical value $\gamma_c = 1.5$, as highlighted by the shaded region.
The wave-packet simulations used a system size $L=800$, with wave-packet parameters $\sigma=40$ and $j_0 = -200$.
$R_L$ and $T_L$ were calculated at time $t=240$.
}
\label{fig:01_time_independent_results}
\end{figure}

Figure~\ref{fig:01_time_independent_results} shows the reflection and transmission coefficients ($R_L$ and $T_L$) for a left-incident wave as a function of $\gamma_1$, with fixed parameters $\gamma_0 = 1$ and $k = \pi/3$. 
The solid lines represent the results from the time-independent scattering method.
Both $R_L$ and $T_L$ exhibit nonmonotonic behavior with increasing $\gamma_1$: they initially increase, reach maxima, and subsequently decrease toward zero. 
This behavior contradicts the intuitive expectation that a larger gain parameter $\gamma_1$ should enhance wave amplification and thus increase $R_L$ and $T_L$.
The unexpected decay of $R_L$ and $T_L$ to zero at large $\gamma_1$ calls into question whether these stationary solutions accurately describe the physical scattering process, motivating further investigation using direct time-dependent wave-packet simulations.

\subsection{Time-dependent simulations of wave-packet evolution}
\label{sec:Time_dependent_method}

To investigate a more physically realistic scenario, we now perform time-dependent simulations of wave-packet evolution on a finite lattice. 
The time evolution of a state $|\Psi(t)\rangle$ is governed by the time-dependent Schrödinger equation
\begin{equation}
i\frac{d}{dt}|\Psi(t)\rangle = H |\Psi(t)\rangle,
\end{equation}
where $H$ is the Hamiltonian of a finite system of size $L$. 
This system comprises the two-site scattering center attached to two truncated leads.
The formal solution is
\begin{equation}
|\Psi(t)\rangle = e^{-iHt}|\Psi(0)\rangle,
\label{eq:time_evolution_wave_function}
\end{equation}
where $|\Psi(0)\rangle$ represents the initial state at $t=0$.

The standard way to solve for $|\Psi(t)\rangle$ is to expand the initial packet $|\Psi(0)\rangle$ in terms of the Hamiltonian's eigenstates. 
However, for a non-Hermitian system, we must consider both the left and right eigenstates of the Hamiltonian.
Let $|\psi_n\rangle$ and $|\phi_n\rangle$ denote the right and left eigenstates of $H$, respectively, which satisfy
\begin{equation}
H|\psi_n\rangle = E_n|\psi_n\rangle, \quad H^\dagger|\phi_n\rangle = E_n^*|\phi_n\rangle,
\label{eq:eigen_equation}
\end{equation}
where $E_n$ represents the complex eigenvalue. 
The eigenstates are chosen to satisfy the biorthogonality condition $\langle \phi_n | \psi_m \rangle = \delta_{n,m}$~\cite{brody2014}.
Assuming the system is not at an exceptional point, we have the completeness relation
\begin{equation}
\sum_n |\psi_n\rangle\langle\phi_n| = 1.
\end{equation}
Expanding the initial state in this basis, we have
\begin{equation}
|\Psi (0)\rangle =\sum _{n} c_{n}(0) |\psi _{n} \rangle,
\end{equation}
where $c_{n}(0) =\langle \phi _{n} |\Psi (0)\rangle$ represents the expansion coefficient at $t=0$. 
Substituting this expansion into Eq.~(\ref{eq:time_evolution_wave_function}), we obtain the time-evolved wave function
\begin{equation}
|\Psi (t)\rangle =\sum _{n} c_{n}( 0) e^{-iE_{n} t} |\psi _{n} \rangle .
\label{eq:wave_function_time_evolution_eigenstates}
\end{equation}

To simulate the scattering process, we initialize the system with a Gaussian wave packet localized in the left lead
\begin{equation}
|\Psi(0)\rangle = \sum_j \Psi_j(0) |j\rangle = \mathcal{N}^{-1} \sum_j e^{-(j-j_0)^2/2\sigma^2} e^{ikj} |j\rangle,
\end{equation}
where $\mathcal{N}$ is a normalization constant.
Our simulations use a wave packet initially centered at $j_0=-200$, with half-width $\sigma=40$ and central wave number $k =\pi/3$.
We employ a finite lattice of 800 sites with truncated leads, consisting of a two-site scattering center ($j=0,1$) and two leads of 399 sites each.
Using Eq.~(\ref{eq:wave_function_time_evolution_eigenstates}), we compute the time-evolved wave function $|\Psi(t)\rangle = \sum_j \Psi_j(t) |j\rangle$.
After the wave packet has fully interacted with the scattering center, we extract the reflection and transmission coefficients from its components in the left and right leads
\begin{equation}
R_{L}= \sum _{j\leq -1} |\Psi _{j}( t) |^{2},
\quad
T_{L}= \sum _{j\geq 2} |\Psi _{j}( t) |^{2},
\label{eq:wave_packet_R_and_T}
\end{equation}
where $t$ is a sufficiently large time such that the reflected and transmitted parts are well separated.

\begin{figure}
\centering
\includegraphics[width=\columnwidth]{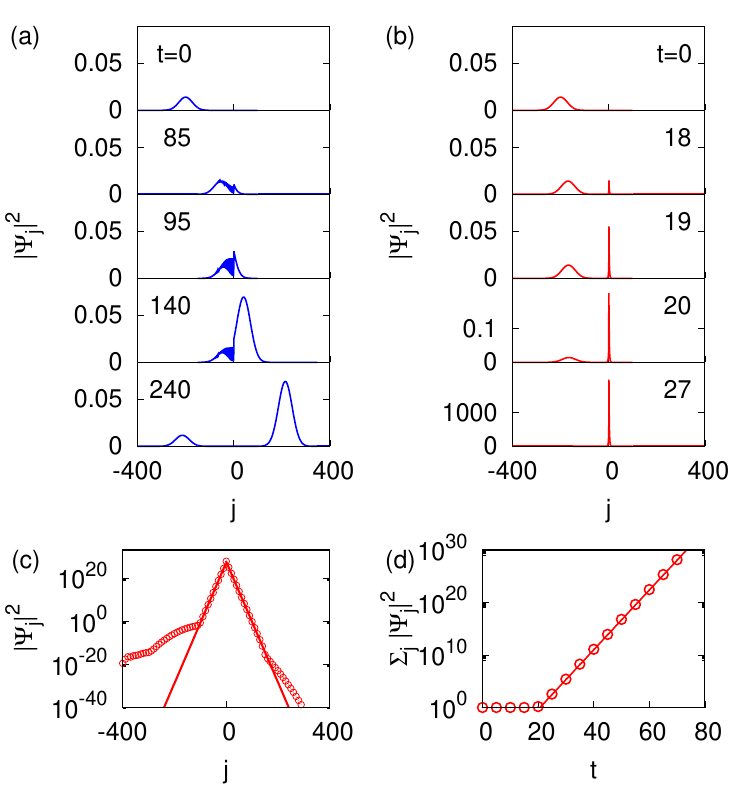}
\caption{
Time evolution of the wave packet for (a) $\gamma_1 = 1.2$, below the critical value $\gamma_c$, and (b) $\gamma_1 = 1.8$, above $\gamma_c$.
(c) Semilogarithmic plot of the intensity profile $|\Psi_j|^2$ at $t=70$ for $\gamma_1 = 1.8$.
The intensity in the central region exhibits an exponential decay, following $C_1 e^{-2 \alpha |j|}$ with $\alpha=0.322 \pm 0.001$.
(d) Semilogarithmic plot of the total intensity $\sum_{j} |\Psi_j|^2$ versus time for $\gamma_1 = 1.8$. 
The total intensity grows exponentially over time, following $C_2 e^{2 \Gamma t}$ with $\Gamma = 0.655 \pm 0.001$. 
These simulations were performed using a system with $\gamma_0=1$, $L=800$, and an initial Gaussian wave packet characterized by $k =\pi/3$, $\sigma=40$, and $j_0 = -200$.
}
\label{fig:02_time_dependent_results} 
\end{figure}

Figure~\ref{fig:02_time_dependent_results}(a) illustrates the wave-packet evolution for $\gamma_0=1$ and $\gamma_1 = 1.2$.
The wave packet propagates towards the scattering center, interacts with it, and is partially reflected and transmitted.
At $t=240$, the calculated reflection and transmission coefficients are $R_L = 0.843$ and $T_L=4.98$, respectively.
These values, plotted as circles and squares in Fig.~\ref{fig:01_time_independent_results}, demonstrate excellent agreement with the time-independent scattering calculations (solid lines).
Further simulations across various $\gamma_1$ values reveal consistent agreement with the time-independent scattering method up to a critical value of $\gamma_c = 1.5$, as illustrated in Fig.~\ref{fig:01_time_independent_results}.

However, a significant discrepancy emerges when $\gamma_1 > 1.5$.
While the time-independent scattering method predicts decreasing reflection and transmission coefficients, time-dependent simulations reveal drastically different behavior.
Figure~\ref{fig:02_time_dependent_results}(b) illustrates the wave-packet evolution for $\gamma_1 = 1.8$.
In contrast to the case with $\gamma_1 = 1.2$, a pronounced peak emerges at the system's center before the wave packet reaches the scattering region, appearing as early as $t = 18$.
This peak grows rapidly and becomes the dominant feature of the wave function at $t = 27$.

To quantify this growth, Fig.~\ref{fig:02_time_dependent_results}(c) displays a semilogarithmic plot of the intensity profile $|\Psi_j|^2$ at $t=70$.
The intensity exhibits exponential decay away from the center, following $C_1 e^{-2 \alpha |j|}$ with $\alpha = 0.322 \pm 0.001$.
Furthermore, Fig.~\ref{fig:02_time_dependent_results}(d) presents a semilogarithmic plot of the total intensity $\sum_{j} |\Psi_j|^2$ versus time.
The total intensity demonstrates exponential growth according to $C_2 e^{2 \Gamma t}$, where $\Gamma = 0.655 \pm 0.001$.
This behavior indicates the emergence of a time-growing bound state within the system. 
This state dominates the system's long-time dynamics, invalidating the conventional scattering picture in which an incident wave packet evolves solely into propagating reflected and transmitted components.
Consequently, scattering coefficients calculated from time-independent scattering methods become unphysical in this regime.

\subsection{Eigenvalues and eigenstates of the finite system}
\label{sec:Eigenvalues_and_eigenstates_of_the_finite_system}

To understand the divergent behavior observed in the time-dependent simulations, we analyze the eigenvalues and eigenstates of the finite system's Hamiltonian.

\begin{figure}
\centering
\includegraphics[width=\columnwidth]{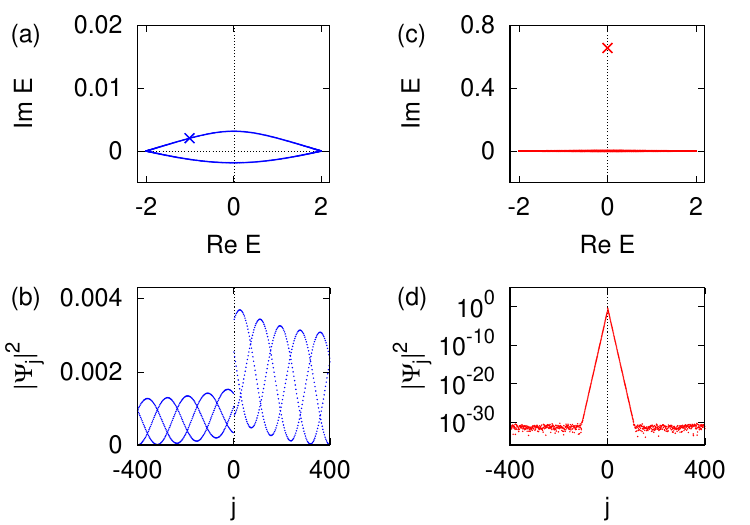}
\caption{
Eigenvalues and eigenstates of the finite system.
(a) Complex energy spectrum for $\gamma_1 = 1.2$ (below the critical value $\gamma_c$), showing eigenvalues clustered near the real axis.
(b) A representative eigenstate from this spectrum [marked by the blue cross in (a)] is extended throughout the system.
(c) Complex energy spectrum for $\gamma_1 = 1.8$ (above $\gamma_c$), where an isolated eigenvalue with a large positive imaginary part ($E \approx 0.655i$) emerges.
(d) The corresponding eigenstate [marked by the red cross in (c)] is a bound state, strongly localized at the scattering center. 
Its intensity profile follows an exponential decay $e^{-2 \alpha |j|}$, with a fitted decay constant $\alpha=0.322 \pm 0.001$.
The calculations were performed for a system with $\gamma_0=1$ and $L=800$.
}
\label{fig:Eigenvalues_and_eigenstates} 
\end{figure}

Figure~\ref{fig:Eigenvalues_and_eigenstates}(a) presents the eigenvalues for a system with size $L=800$, $\gamma_0=1$ and $\gamma_1 = 1.2$, which is below the critical value $\gamma_c$.
The real parts of the eigenenergies span from $-2$ to 2 and possess small imaginary parts, forming a spectrum clustered around the real axis.
The associated wave functions are extended throughout the system, as depicted in Fig.~\ref{fig:Eigenvalues_and_eigenstates}(b).
These states are the finite-size counterparts of the real-energy continuous scattering states of the infinite system.
Crucially, their imaginary energy components arise from finite-size effects, scaling as $1/L$ and vanishing in the infinite-system limit.
Although these states have positive imaginary energy parts, they do not cause divergence during the scattering process, as shown in Fig.~\ref{fig:02_time_dependent_results}(a)\footnote{
We note that while these scattering states with small imaginary parts do not cause divergence during the scattering process, they can nevertheless lead to exponential intensity growth in a finite system over long timescales.
This occurs because a wave packet, after the initial scattering event, repeatedly reflects off the system boundaries and rescatters off the central region, leading to cumulative amplification.
This process is a finite-size artifact and is distinct from the intrinsic time-growing bound state discussed next.
}.

The situation changes dramatically when $\gamma_1 = 1.8$, which exceeds $\gamma_c$.
As shown in Fig.~\ref{fig:Eigenvalues_and_eigenstates}(c), an isolated discrete eigenvalue with a significant positive imaginary part emerges from the continuum.
The corresponding eigenstate is strongly localized at the scattering center [Fig.~\ref{fig:Eigenvalues_and_eigenstates}(d)], confirming its nature as a time-growing bound state. 
The imaginary part of its energy, $\operatorname{Im}(E) \approx 0.655$, and the spatial decay rate of its wave function, determined to be $\alpha=0.322 \pm 0.001$, are in excellent agreement with the the temporal growth rate $\Gamma=0.655 \pm 0.001$ and spatial decay rate $\alpha=0.322 \pm 0.001$ observed in the wave-packet simulation [Figs.~\ref{fig:02_time_dependent_results}(c) and (d)]. 
This provides conclusive evidence that the excitation of this discrete bound state is responsible for the system's divergent long-time behavior.
Crucially, this imaginary energy component is a robust feature that persists in the infinite-system limit.
It is the emergence of this specific, exponentially growing bound state that causes the system dynamics to diverge rapidly [as seen in Fig.~\ref{fig:02_time_dependent_results}(b)] and invalidates the conventional scattering picture.

Therefore, the mere presence of complex energies in a finite system is not a sufficient condition for concluding that the actual wave-packet dynamics will diverge from the predictions of time-independent scattering calculations.
The divergence specifically arises from the emergence of time-growing bound states, characterized by eigenvalues with nonvanishing positive imaginary parts in the infinite-system limit.

\subsection{$S$-matrix poles}
\label{sec:S_matrix_poles}

While the finite-system eigenvalue analysis in Sec.~\ref{sec:Eigenvalues_and_eigenstates_of_the_finite_system} provides strong numerical evidence for the emergence of a time-growing bound state, a more rigorous and analytical approach is to study the $S$ matrix of the infinite system. 
The poles of the $S$ matrix in the complex $k$ or energy plane provide a precise system-size-independent signature of the system's bound and resonant states~\cite{moiseyev2011,sasada2011,hatano2014,garmon2015}. 
This method allows us to analytically determine the exact conditions under which time-growing bound states emerge.

The $S$ matrix, which encapsulates the full scattering information, is defined as
\begin{equation}
S= \begin{pmatrix} r_L & t_R \\ t_L & r_R \end{pmatrix}.
\end{equation} 
The $S$-matrix poles occur when the denominators of the scattering amplitudes ($r_L$, $t_L$, $r_R$, and $t_R$) vanish. 
Based on our previous derivations in Eqs.~(\ref{eq:rL}), (\ref{eq:tL}), (\ref{eq:rR}), and (\ref{eq:tR}), this condition yields:
\begin{equation}
\gamma _{0} -\gamma _{1} +i\gamma _{0} \gamma _{1} e^{ik} +2\sin k = 0.
\label{eq:pole_equation}
\end{equation}

These poles can also be understood by expressing the scattering amplitudes as $r_L = B/A$, $t_L = C/A$, $r_R = C/D$, and $t_R = B/D$, where $A$, $B$, $C$, and $D$ are the coefficients of the scattering eigenstates [Eqs.~(\ref{eq:wave_function_left_scattering}) and (\ref{eq:wave_function_right_scattering})].
When either $A$ or $D$ equals zero, a pole emerges, resulting in eigenstates of the form:
\begin{equation}
\psi _{S}^{k} (j)=\begin{cases}
Be^{-ikj} & \text{for } j\leq -1,\\
\psi _{S}^{k} (0) & \text{for } j=0,\\
\psi _{S}^{k} (1) & \text{for } j=1,\\
Ce^{ikj} & \text{for } j\geq 2.
\end{cases}
\end{equation}
This represents a purely outgoing wave at both ends of the system, a condition known as the Siegert boundary condition~\cite{siegert1939}.
When this boundary condition is applied to the Schrödinger equations (\ref{eq:Schrodinger_equation_leads})--(\ref{eq:Schrodinger_equation_scattering_center_site1}),  it yields the same pole equation as Eq.~(\ref{eq:pole_equation}). 
In contrast to scattering states where the wave number $k$ takes continuous values from 0 to $\pi$, the Siegert boundary condition produces a discrete set of complex wave numbers $k_n$ with associated complex energies $E_n = -2\cos k_n$.

The time evolution of the eigenstate associated with the pole $k_n$ is given by
\begin{equation}
\begin{aligned}
\psi _{S}^{k} (j,t) & =e^{-iE_{n} t} \psi _{S}^{k} (j)\\
 & =e^{-iE_{n}^{r} t} e^{E_{n}^{i} t} \times \begin{cases}
Be^{-ik_{n}^{r} j} e^{k_{n}^{i} j} & \text{for } j\leq -1,\\
\psi _{S}^{k} (0) & \text{for } j=0,\\
\psi _{S}^{k} (1) & \text{for } j=1,\\
Ce^{ik_{n}^{r} j} e^{-k_{n}^{i} j} & \text{for } j\geq 2.
\end{cases}
\end{aligned}
\end{equation}
Here, $E_n^r$ and $E_n^i$ represent the real and imaginary parts of $E_n$, while $k_n^r$ and $k_n^i$ denote the corresponding parts of $k_n$. 
Each component has a distinct physical interpretation.
The real part of the wave number $k_n^r$ determines the wave propagation direction: positive values indicate outward propagation from the scattering center, while negative values signify incoming waves. 
The imaginary part $k_n^i$ governs the spatial behavior: positive values yield exponential decay as $|j| \to \infty$ (characteristic of bound states), while negative values lead to exponential growth.
The imaginary part of the energy $E_n^i$ controls the temporal evolution: positive values produce exponential growth over time, while negative values result in exponential decay. 
Given the relation $E_n = -2\cos k_n$, we have $E_{n}^{i} =2\sin k_{n}^{r}\sinh k_{n}^{i}$.

The nature of these discrete states is determined by the location of $k$ in the complex plane~\cite{sasada2011,hatano2014,garmon2015}. 
In the first quadrant where $k_n^r > 0$ and $k_n^i > 0$, we find $E_n^i > 0$, corresponding to time-growing outgoing bound states. 
In the second quadrant where $k_n^r < 0$ and $k_n^i > 0$, we have $E_n^i < 0$, representing time-decaying incoming bound states. 
In the third quadrant where $k_n^r < 0$ and $k_n^i < 0$, $E_n^i > 0$ indicates time-growing incoming antiresonant states. 
In the fourth quadrant where $k_n^r > 0$ and $k_n^i < 0$, we obtain $E_n^i < 0$, describing time-decaying outgoing resonant states.
Among these states, only the bound states in the first and second quadrants are normalizable and lie within the Hilbert space of the system.
While spatially divergent, the resonant states still play a crucial role in determining the system's scattering properties.

We now proceed to solve Eq.~(\ref{eq:pole_equation}) to determine the $S$-matrix poles.
Using the same parameters as in Figs.~\ref{fig:01_time_independent_results}, \ref{fig:02_time_dependent_results} and \ref{fig:Eigenvalues_and_eigenstates}, we set $\gamma_0=1$ and analyze how the poles evolve with varying $\gamma_1$.
Equation~(\ref{eq:pole_equation}) then simplifies to
\begin{equation}
1-\gamma_1 + i\gamma_1 e^{ik} + 2\sin k = 0.
\end{equation}
Introducing the substitution $z = e^{ik}$ transforms this into a quadratic equation
\begin{equation}
z^2(\gamma_1 - 1) + i(\gamma_1 - 1)z + 1 = 0.
\end{equation}
The solutions are
\begin{equation}
z_{1,2} = \frac{1}{2}\left(-i \pm \sqrt{\frac{\gamma_1 + 3}{1-\gamma_1}}\right).
\end{equation}
The corresponding values of $k$ can be obtained through $k_{1,2} = -i\ln z_{1,2}$.
For $0 < \gamma_1 < 1$, we find
\begin{gather}
k_1 = -\pi + \arcsin\left(\frac{\sqrt{1-\gamma_1}}{2}\right) + i\frac{1}{2}\ln(1-\gamma_1)\label{eq:k1_less_than_1},\\
k_2 = -\arcsin\left(\frac{\sqrt{1-\gamma_1}}{2}\right) + i\frac{1}{2}\ln(1-\gamma_1).\label{eq:k2_less_than_1}
\end{gather}
For $\gamma_1 > 1$, the solutions become
\begin{gather}
k_{1}  =-\frac{\pi }{2} -i\ln\left[\frac{1}{2}\left( 1+\sqrt{\frac{\gamma _{1} +3}{\gamma _{1} -1}}\right)\right],\\
k_{2}  =\frac{\pi }{2} -i\ln\left[\frac{1}{2}\left( -1+\sqrt{\frac{\gamma _{1} +3}{\gamma _{1} -1}}\right)\right].
\end{gather}
Here, we restrict $-\pi < \operatorname{Re} k\leq \pi $. 

\begin{figure}
\centering
\includegraphics[width=\columnwidth]{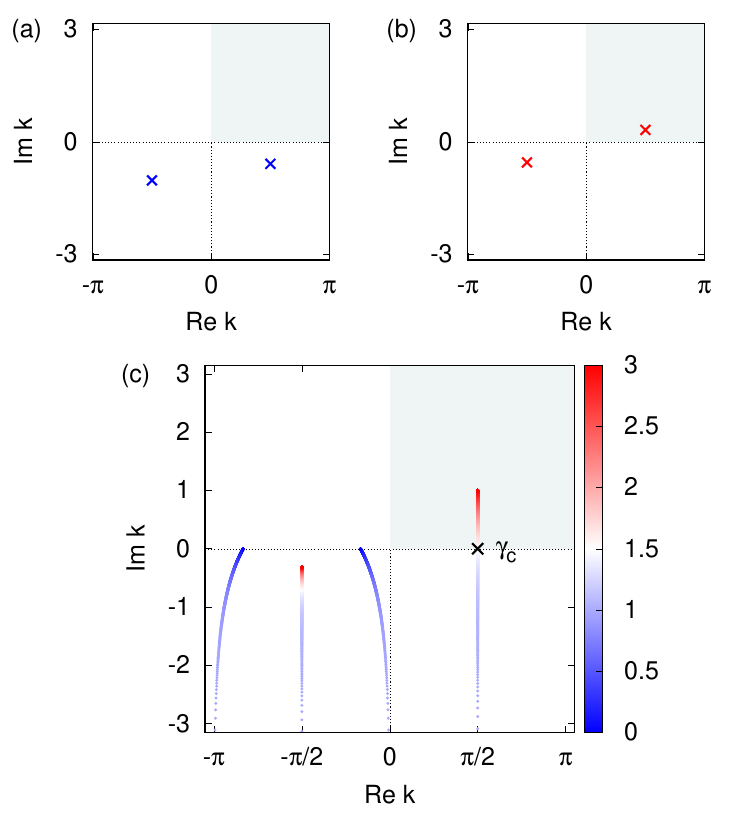}
\caption{
Locations of $S$-matrix poles in the complex $k$ plane for (a) $\gamma_1 = 1.2$, below the critical value $\gamma_c$, and (b) $\gamma_1 = 1.8$, above $\gamma_c$.
(c) Trajectories of the $S$-matrix poles as a function of $\gamma_1$.
The color bar shows the value of $\gamma_1$.
A pole crosses the real axis and enters the first quadrant (shaded region) at the critical value $\gamma_c=1.5$, indicating the emergence of a time-growing bound state in the system.
For these calculations, $\gamma_0$ was fixed at 1.
}
\label{fig:03_distribution_of_poles_1}
\end{figure}

Figures~\ref{fig:03_distribution_of_poles_1}(a) and (b) illustrate the pole distributions in the complex $k$ plane for $\gamma_1 = 1.2$ and 1.8, respectively.
When $\gamma_1 = 1.2$, both poles are located outside the first quadrant, indicating the absence of time-growing bound states. 
Consequently, the time-independent scattering calculation provides a physically relevant description of the wave-packet dynamics in this regime, consistent with our previous observations in Figs.~\ref{fig:01_time_independent_results} and \ref{fig:02_time_dependent_results}.
However, for $\gamma_1 = 1.8$, pole $k_2$ moves into the first quadrant ($k_2 \approx 1.5708 + 0.322i$), yielding a complex energy of $E_2 \approx 0.655i$.
The corresponding eigenstate, given by $\psi_S^k(j,t) \propto e^{E_2^i t}e^{-k_2^i|j|}$, demonstrates both spatial localization and temporal growth.
This result aligns precisely with our time-dependent simulation results in Figs.~\ref{fig:02_time_dependent_results}(c) and (d), where we observed exponential spatial decay with rate $\alpha = 0.322$ and temporal growth with rate $\Gamma = 0.655$.

Figure~\ref{fig:03_distribution_of_poles_1}(c) shows the complete trajectories of the poles as $\gamma_1$ varies.
Initially, both poles lie on the negative real axis of the complex $k$ plane.
As $\gamma_1$ increases, they move downward and become antiresonant states.
When $\gamma_1$ approaches 1, both poles move towards $-i\infty $ [see Eqs.~(\ref{eq:k1_less_than_1}) and (\ref{eq:k2_less_than_1})].
For $\gamma_1 > 1$, while pole $k_1$ remains an antiresonant state, $k_2$ becomes a resonant state, and both poles gradually move upward with increasing $\gamma_1$.
The critical point $\gamma_c$ occurs when pole $k_2$ crosses the positive real axis, satisfying:
\begin{equation}
\ln\left[\frac{1}{2}\left( -1+\sqrt{\frac{\gamma _{1} +3}{\gamma _{1} -1}}\right)\right] =0.
\end{equation}
This equation yields 
\begin{equation}
\gamma_c = 1.5,
\end{equation}
which coincides with the critical value observed in Fig.~\ref{fig:01_time_independent_results}, beyond which the actual wave-packet dynamics diverge dramatically from the predictions of time-independent scattering calculations.
This crossing point on the real axis represents a spectral singularity~\cite{mostafazadeh2009,longhi2009,ramezani2014,wang2016,jin2018b}.
Beyond $\gamma_c$, $k_2$ transitions from a resonant state to a time-growing bound state.
As $\gamma_1$ increases further, $k_1$ approaches $-\pi/2$, while $k_2$ rises indefinitely.

Notably, in Hermitian systems, poles cannot enter the first or second quadrant of the complex $k$ plane. 
This phenomenon is unique to non-Hermitian systems.
Although such pole behavior has been previously reported~\cite{garmon2015,shobe2021}, its implications for the physical relevance of time-independent scattering calculations were not fully recognized.
This oversight can lead to misleading conclusions about the actual dynamics of the wave packet.

\subsection{Physical relevance of time-independent scattering calculations in non-Hermitian systems}
\label{sec:Physical_relevance_of_time_independent_scattering_calculations_in_non_Hermitian_systems}

We now explain why the reflection and transmission coefficients derived from the time-independent scattering method become physically irrelevant to the actual evolution of a wave packet in certain parameter regimes.
The method, as presented in Sec.~\ref{sec:Time_independent_method}, rests on the fundamental assumption that the incident wave is a pure superposition of the system's scattering states.
The reflection and transmission coefficients are defined within this framework for these stationary scattering states.
However, any physically realistic incident wave packet, such as the Gaussian profile used in Sec.~\ref{sec:Time_dependent_method}, will generally have a nonzero overlap with the evanescent tails of the system's bound states, even when prepared far from the scattering region.
This overlap, however small, violates the basic premise of the time-independent scattering method and can lead to a stark divergence between the predicted and actual dynamics.

A rigorous description of the wave packet's evolution must account for the complete set of the Hamiltonian's eigenstates, which includes both the continuous scattering states $|\psi_{L/R}^k\rangle$ and the discrete bound states $|\psi_b\rangle$~\cite{hammer1977,stvneng1991}. 
These eigenstates form a complete basis, satisfying the completeness relation
\begin{equation}
\int_0^\pi dk(|\psi_L^k\rangle\langle\phi_L^k| + |\psi_R^k\rangle\langle\phi_R^k|) + \sum_b |\psi_b\rangle\langle\phi_b| = 1,
\end{equation}
where $|\phi_{L,R}^k\rangle$ and $|\phi_b\rangle$ denote the corresponding left eigenstates of the Hamiltonian, as defined in Eq.~(\ref{eq:eigen_equation}).
The initial state $|\Psi(0)\rangle$ can therefore be expanded as
\begin{equation}
|\Psi (0)\rangle =\int _{0}^{\pi } dk\left[ c_{L}^{k}( 0) |\psi _{L}^{k} \rangle +c_{R}^{k}( 0) |\psi _{R}^{k} \rangle \right] +\sum _{b} c_{b}( 0) |\psi _{b} \rangle ,
\end{equation}
with expansion coefficients
\begin{equation}
\begin{aligned}
c_L^k(0) &= \langle\phi_L^k|\Psi(0)\rangle, \\
c_R^k(0) &= \langle\phi_R^k|\Psi(0)\rangle, \\
c_b(0) &= \langle\phi_b|\Psi(0)\rangle.
\end{aligned}
\end{equation}
The time evolution then follows
\begin{equation}
|\Psi (t)\rangle =\int _{0}^{\pi } dk\left[ c_{L}^{k} (t)|\psi _{L}^{k} \rangle +c_{R}^{k} (t)|\psi _{R}^{k} \rangle \right] +\sum _{b} c_{b} (t)|\psi _{b} \rangle ,
\end{equation}
where the time-dependent coefficients are
\begin{equation}
\begin{aligned}
c_L^k(t) &= c_L^k(0)e^{-iE_k t}, \\
c_R^k(t) &= c_R^k(0)e^{-iE_k t}, \\
c_b(t) &= c_b(0)e^{-iE_b t}.
\end{aligned}
\end{equation}
Here, $E_k = -2\cos k$ is the real-valued energy of the scattering states, while $E_b$ denotes the bound-state energies, which can be complex in non-Hermitian systems. 
Since bound states are exponentially localized, the initial overlap $c_b(0)$ is typically negligible for a wave packet prepared far from the scattering region.

In Hermitian systems, where $E_b$ is real, the term $e^{-iE_b t}$ is a pure phase factor. 
The magnitude of the bound-state contribution, $|c_b(t)|=|c_b(0)|$, remains constant and small, rendering the violation of the pure-scattering-state assumption inconsequential. 
Consequently, the system's dynamics is dominated by the scattering components, and the time-independent scattering calculation accurately predicts the asymptotic behavior of the scattered wave packet.

In non-Hermitian systems, however, the situation can be dramatically different.
The bound-state energies are generally complex, $E_b = E_b^r + iE_b^i$. 
The corresponding coefficient evolves as
\begin{equation}
c_b(t) = c_b(0)e^{-iE_b^r t}e^{E_b^i t}.
\end{equation}
When $E_b^i < 0$, the bound states decay exponentially with time and can be safely neglected.
Conversely, when $E_b^i > 0$, these states exhibit exponential growth.
Despite a small initial overlap $c_b(0)$, this exponential growth eventually dominates the system's dynamics, as demonstrated in Fig.~\ref{fig:02_time_dependent_results}(b). 
As a result, the long-time dynamics is governed by this nonpropagating, growing state rather than by the propagating waves. 
This invalidates the conventional scattering picture, rendering the transmission and reflection coefficients derived from the time-independent scattering method physically irrelevant to the system's long-time evolution.
This explains the discrepancy observed in Fig.~\ref{fig:01_time_independent_results}. 
While the coefficients $R$ and $T$ derived from the time-independent scattering method correctly describe the scattering properties of the scattering-state components of the wave packet, the coefficients extracted numerically from the wave-packet simulation [Eq.~(\ref{eq:wave_packet_R_and_T})] capture the total probability in the leads, which includes contributions from both the scattering and bound-state components. 
In the presence of such a growing state, the very concept of a time-independent asymptotic scattering coefficient becomes ill-defined.

A more rigorous definition of scattering coefficients can be formulated using the probability current,
\begin{equation}
	\mathcal{J}(j,t) = i[\psi^*(j+1,t)\psi(j,t) - \psi(j+1,t)\psi^*(j,t)].
\end{equation}
For a stationary scattering state of the form
\begin{equation}
\psi _{L}^{k} (j)=\begin{cases}
Ae^{ikj} +Be^{-ikj} & \text{for } j\leq -1,\\
Ce^{ikj} & \text{for } j\geq 2,
\end{cases}
\end{equation}
the probability current is found to be
\begin{equation}
	\mathcal{J}=2\sin k\times \begin{cases}
|A|^{2} -|B|^{2} & \text{for } j\leq -1,\\
|C|^{2} & \text{for } j\geq 2.
\end{cases}
\end{equation}
As expected for stationary scattering states, the current is constant in both time and position within the leads.
The reflection and transmission coefficients are then defined as the ratios of the reflected and transmitted currents to the incident current.
This yields $R_L = |\mathcal{J}_R|/\mathcal{J}_I = |B|^2/|A|^2$ and $T_L = \mathcal{J}_T/\mathcal{J}_I = |C|^2/|A|^2$, which are identical to the definitions used in Sec.~\ref{sec:Time_independent_method}.
In sharp contrast, for a bound state
\begin{equation}
\psi _{b} (j,t)=e^{-iE_{b}^{r} t} e^{E_{b}^{i} t} \times \begin{cases}
Be^{-ik_{b}^{r} j} e^{k_{b}^{i} j} & \text{for } j\leq -1,\\
Ce^{ik_{b}^{r} j} e^{-k_{b}^{i} j} & \text{for } j\geq 2,
\end{cases}
\end{equation}
the calculated probability current is
\begin{equation}
\mathcal{J} (j,t)=2\sin (k_{b}^{r} )e^{2E_{b}^{i} t} \times \begin{cases}
-|B|^{2} e^{k_{b}^{i} (2j+1)} & \text{for } j\leq -1,\\
|C|^{2} e^{-k_{b}^{i} (2j+1)} & \text{for } j\geq 2.
\end{cases}
\end{equation}
Crucially, this current is neither spatially uniform nor temporally constant.
For $E_b^i>0$, the current at any given position grows exponentially in time.
Consequently, the total current arising from a physical wave packet, which includes this bound-state component, will be dominated by this exponentially growing contribution.
This dominance ultimately results in divergent reflection and transmission coefficients in the long-time limit.

\section{Asymmetric hopping models}
\label{sec:Asymmetric_hopping_models}

In addition to the imaginary on-site potentials discussed in Sec.~\ref{sec:Imaginary_on_site_potential_Model}, non-Hermiticity can also arise from asymmetric hopping terms. 
These models have been extensively studied in various contexts, including anti-$\mathcal{PT}$-symmetric systems~\cite{peng2016,choi2018,jin2018a,xu2021,xu2023a}, non-Hermitian topological systems~\cite{yao2018,kunst2018,gong2018,lieu2018,liu2019a}, and non-Hermitian disordered systems~\cite{liu2021a,schiffer2021,tang2021,kawabata2021,molignini2023,li2024,halder2025}.
We demonstrate that time-growing bound states can emerge in these systems as well, potentially rendering time-independent scattering calculations physically irrelevant to the actual wave-packet dynamics.
Therefore, when studying transport properties using time-independent scattering methods, careful examination of $S$-matrix pole distributions is essential.

\begin{figure*}
\centering
\includegraphics[width=2\columnwidth]{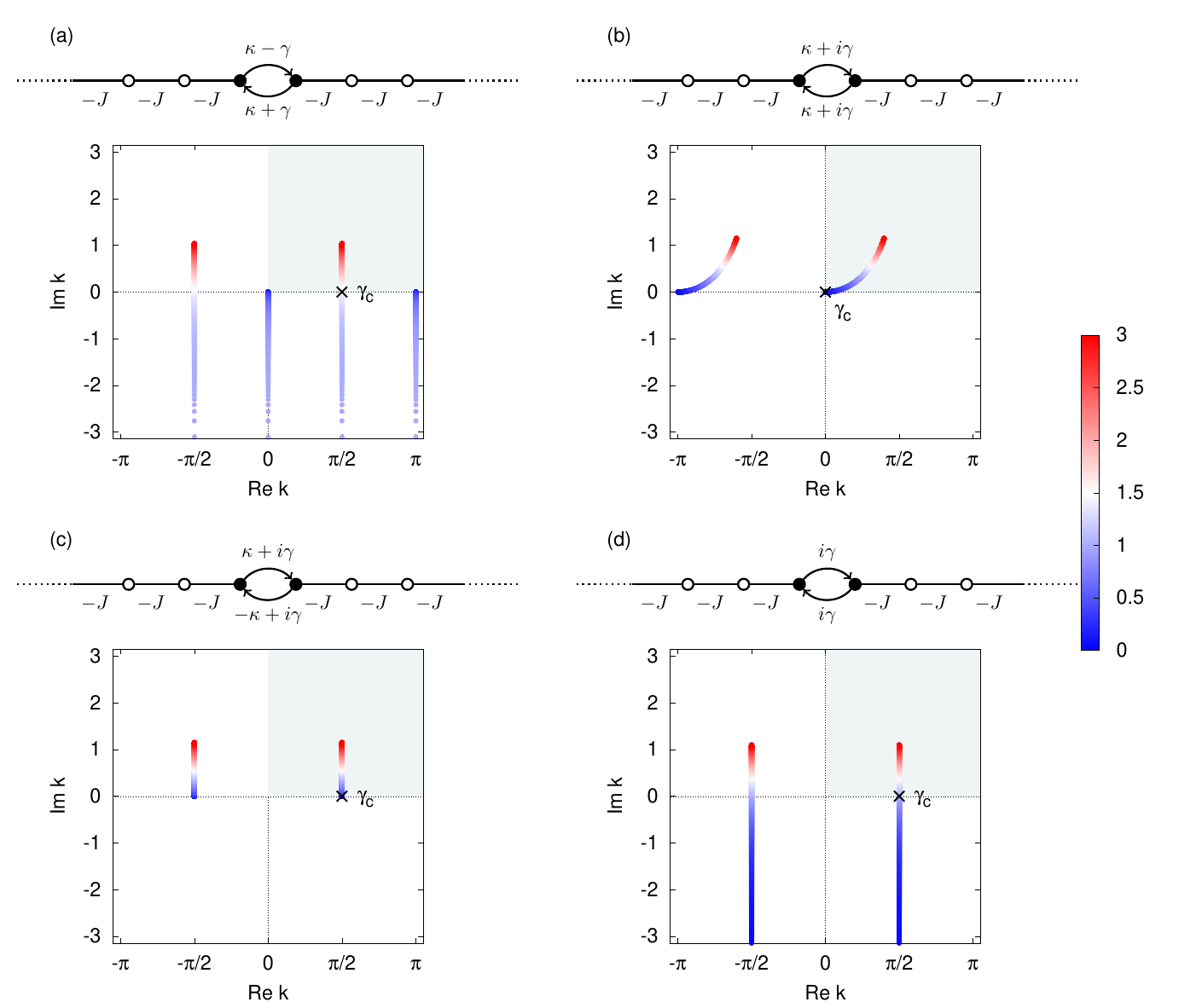}
\caption{
Evolution of $S$-matrix poles in the complex $k$ plane as a function of $\gamma$ for different asymmetric hopping models, as schematically illustrated at the top of each panel.
The color bar shows the value of $\gamma$.
These models are characterized as follows: (a) unequal hopping with $\kappa_R=\kappa-\gamma$ and $\kappa_L=\kappa+\gamma$, 
(b) complex hopping where $\kappa_R=\kappa_L=\kappa+i\gamma$, 
(c) anti-Hermitian hopping with $\kappa_R=\kappa+i\gamma$ and $\kappa_L=-\kappa+i\gamma$, 
(d) imaginary coupling where $\kappa_R=\kappa_L=i\gamma$.
In all cases, $\kappa=-1$. 
The corresponding critical values $\gamma_c$ are (a) $\sqrt{2}$, (b) 0, (c) 0, and (d) 1.
}
\label{fig:04_distribution_of_poles_2_combined}
\end{figure*}

We consider a system similar to that in Sec.~\ref{sec:Imaginary_on_site_potential_Model}, consisting of a scattering center connected to two semi-infinite leads. 
The scattering center Hamiltonian $H_C$ now features asymmetric hopping
\begin{equation}
H_{C} =\kappa _{L} |0\rangle \langle 1|+\kappa _{R} |1\rangle \langle 0|,
\end{equation}
where $\kappa _{L}$ and $\kappa _{R}$ represent the hopping amplitudes from right to left and from left to right, respectively. 
The system becomes non-Hermitian when $\kappa _{L} \neq \kappa _{R}^{*}$. 
We examine four distinct types of asymmetric hopping, as shown in Fig.~\ref{fig:04_distribution_of_poles_2_combined}.
The unequal hopping case has $\kappa_R = \kappa - \gamma$ and $\kappa_L = \kappa + \gamma$. 
For complex hopping, both amplitudes are equal with $\kappa_R = \kappa_L = \kappa + i\gamma$. 
The anti-Hermitian hopping model has $\kappa_R = \kappa + i\gamma$ and $\kappa_L = -\kappa + i\gamma$, while the imaginary coupling case features purely imaginary and equal hopping with $\kappa_R = \kappa_L = i\gamma$. 
Here, both $\kappa$ and $\gamma$ are real parameters. 
In the following analysis, we set $\kappa=-1$ and analyze how the $S$-matrix poles evolve with varying $\gamma$. 
The appearance of a pole in the first quadrant of the complex $k$ plane indicates the emergence of a time-growing bound state.
The presence of such a mode dominates the wave-packet dynamics, rendering the transmission and reflection coefficients obtained from time-independent scattering methods unphysical.
Since the time-dependent results are similar to those presented in Sec.~\ref{sec:Time_dependent_method}, we omit them here to avoid redundancy.

To analyze the $S$-matrix poles, we first derive their governing equation. 
The Schrödinger equations for the scattering center become
\begin{align}
-\psi (-1)+\kappa _{L} \psi(1) & =E\psi (0),\\
\kappa _{R} \psi (0)-\psi (2) & =E\psi (1),
\end{align}
where $E = -2\cos k$.
Following the same procedure as in Sec.~\ref{sec:Time_independent_method}, we obtain the reflection and transmission amplitudes
\begin{align}
r_{L} &=\frac{(\kappa _{L} \kappa _{R} -1)e^{2ik}}{1-\kappa _{L} \kappa _{R} e^{2ik}},\\
t_{L} &=\frac{\kappa _{R}( e^{2ik} -1)}{1-\kappa _{L} \kappa _{R} e^{2ik}},\\
r_{R} &=\frac{\kappa _{L} \kappa _{R} -1}{1-\kappa _{L} \kappa _{R} e^{2ik}},\\
t_{R} &=\frac{\kappa _{L}( e^{2ik} -1)}{1-\kappa _{L} \kappa _{R} e^{2ik}}.
\end{align}
The $S$-matrix poles are determined by the zeros of the denominator, yielding
\begin{equation}
1-\kappa _{L} \kappa _{R} e^{2ik} =0.
\label{eq:pole_equation_asymmetric}
\end{equation}
We solve this equation within the range $-\pi < \operatorname{Re} k\leq \pi$.

\subsection{Unequal hopping model}
\label{sec:Unequal_hopping_model}

The unequal hopping model features real but asymmetric hopping amplitudes: $\kappa_R = \kappa - \gamma$ and $\kappa_L = \kappa + \gamma$.
Hatano and Nelson introduced this model in 1996~\cite{hatano1996,hatano1997}, demonstrating that nonreciprocal hopping can suppress Anderson localization in 1D systems.
This discovery initiated extensive research into non-Hermitian disordered systems.
The model has subsequently become instrumental in studying various phenomena, including the non-Hermitian skin effect~\cite{lee2016,yao2018,kunst2018,zhang2022a}, non-Hermitian topological phases~\cite{yao2018,kunst2018,gong2018,lieu2018,liu2019a}, and non-Hermitian disordered systems~\cite{liu2021a,schiffer2021,tang2021,kawabata2021,molignini2023,li2024,halder2025}.

Setting $\kappa=-1$, Eq.~(\ref{eq:pole_equation_asymmetric}) reduces to:
\begin{equation}
e^{2ik} =\frac{1}{1-\gamma ^{2}}.
\end{equation}
For $0< \gamma < 1$, the solutions are
\begin{gather}
k_{1}  = i\frac{1}{2}\ln\left( 1-\gamma ^{2}\right),\\
k_{2}  =\pi +i\frac{1}{2}\ln\left( 1-\gamma ^{2}\right).
\end{gather}
For $\gamma > 1$, we have
\begin{gather}
k_{1}  =-\frac{\pi }{2} +i\frac{1}{2}\ln\left( \gamma ^{2} -1\right),\\
k_{2}  =\frac{\pi }{2} +i\frac{1}{2}\ln\left( \gamma ^{2} -1\right).
\end{gather}
Figure~\ref{fig:04_distribution_of_poles_2_combined}(a) shows the pole trajectories as a function of $\gamma$. 
The poles, initially located at 0 and $\pi$, move downward as $\gamma$ increases.
As $\gamma$ approaches 1, both poles tend toward $-i\infty$.  
For $\gamma > 1$, the poles emerge from $\pm \frac{\pi}{2} - i\infty$ and move upward.
The critical point $\gamma_c$ occurs when pole $k_2$ crosses the real axis, satisfying
\begin{equation}
\frac{1}{2}\ln\left( \gamma _{c}^{2} -1\right) =0,
\end{equation}
which yields $\gamma _{c} =\sqrt{2}$. 
Beyond this critical value, a time-growing bound state emerges, rendering the predictions from time-independent scattering calculations a physically irrelevant description of the wave-packet dynamics.

\subsection{Complex hopping model}
\label{sec:Complex_hopping_model}

In the complex hopping model, the hopping amplitudes are equal but complex: $\kappa_R = \kappa_L = \kappa + i\gamma$~\cite{takata2022}. 
The system's non-Hermiticity originates from the imaginary component $\gamma$.

With $\kappa=-1$, the pole equation becomes
\begin{equation}
e^{2ik} =\frac{1}{(i\gamma -1)^{2}}.
\end{equation}
The solutions are
\begin{gather}
k_{1}  =-\pi +\arctan \gamma +i\frac{1}{2}\ln\left( 1+\gamma ^{2}\right),\\
k_{2}  =\arctan \gamma +i\frac{1}{2}\ln\left( 1+\gamma ^{2}\right).
\end{gather}
As shown in Fig.~\ref{fig:04_distribution_of_poles_2_combined}(b), pole $k_2$ enters the first quadrant immediately for any nonzero $\gamma$. 
The critical point occurs at
\begin{equation}
\frac{1}{2}\ln\left( 1+\gamma _{c}^{2}\right) =0,
\end{equation}
which gives $\gamma _{c} =0$, indicating that for any nonzero $\gamma$, time-independent scattering calculations provide a physically irrelevant description of the wave-packet dynamics.

\subsection{Anti-Hermitian hopping model}
\label{sec:Anti_Hermitian_hopping_model}

The anti-Hermitian hopping model features $\kappa _{R} =\kappa +i\gamma$ and $\kappa _{L} =-\kappa +i\gamma$~\cite{li2021}. 
The scattering center exhibits anti-Hermiticity, satisfying $H_{c} =-H_{c}^{\dagger}$.
Furthermore, the system also possesses anti-$\mathcal{PT}$ symmetry, satisfying $H_{\mathrm{c}} =-(\mathcal{PT} )H_{\mathrm{c}} (\mathcal{PT} )^{-1}$~\cite{peng2016,jin2018a,xu2021,xu2023a}. 
Here,  $\mathcal{P}$ represents the parity operator
\begin{equation}
\mathcal{P} = \begin{pmatrix} 0 & 1 \\ 1 & 0 \end{pmatrix}
\end{equation}
and $\mathcal{T}$ denotes the time-reversal operator that performs complex conjugation: $\mathcal{T} i\mathcal{T}^{-1} =-i$.

Setting $\kappa=-1$, the pole equation becomes
\begin{equation}
e^{2ik} =-\frac{1}{\gamma ^{2} +1},
\end{equation}
with solutions
\begin{gather}
k_{1}  =-\frac{\pi }{2} +i\frac{1}{2}\ln\left( 1+\gamma ^{2}\right),\\
k_{2}  =\frac{\pi }{2} +i\frac{1}{2}\ln\left( 1+\gamma ^{2}\right).
\end{gather}
Figure~\ref{fig:04_distribution_of_poles_2_combined}(c) shows that pole $k_2$ enters the first quadrant for any positive $\gamma$.
The critical value $\gamma_c$ is determined by
\begin{equation}
\frac{1}{2}\ln\left( 1+\gamma _{c}^{2}\right) =0,
\end{equation}
which yields $\gamma_c = 0$. 
Consequently, similar to the complex hopping model, time-independent scattering calculations yield a physically irrelevant description of the wave-packet dynamics for any nonzero $\gamma$.

\subsection{Imaginary coupling model}
\label{sec:Imaginary_coupling_model}

The imaginary coupling model is characterized by $\kappa _{R} = \kappa _{L}=i\gamma$, which represents a special case of both complex hopping and anti-Hermitian hopping models when $\kappa=0$.
The scattering center maintains both anti-Hermiticity and anti-$\mathcal{PT}$ symmetry~\cite{peng2016,choi2018,jin2018a,xu2021,xu2023a}.

The pole equation simplifies to
\begin{equation}
e^{2ik} =-\frac{1}{\gamma ^{2}},
\end{equation}
with solutions
\begin{gather}
k_{1}  =-\frac{\pi }{2} +i\ln \gamma,\\
k_{2}  =\frac{\pi }{2} +i\ln \gamma.
\end{gather}
As shown in Fig.~\ref{fig:04_distribution_of_poles_2_combined}(d), at $\gamma=0$, the poles are located at $\pm \frac{\pi }{2}-i\infty$. 
As $\gamma$ increases, both poles move upward. 
The critical point occurs when pole $k_2$ reaches the real axis, corresponding to
\begin{equation}
\ln \gamma _{c} =0,
\end{equation}
giving $\gamma _{c} =1$. 
When $\gamma$ exceeds this critical value, a time-growing bound state emerges, rendering the predictions of time-independent scattering calculations physically irrelevant to the actual wave-packet dynamics.

\section{Discussion}
\label{sec:Discussion}

In conclusion, we have systematically investigated the conditions under which time-independent scattering calculations yield a physically relevant description of wave-packet dynamics in non-Hermitian systems.
Our analysis clarifies a crucial distinction between the formal self-consistency of time-independent scattering methods and their applicability to realistic physical scenarios.
While these methods assume that the incident wave is a pure superposition of scattering states, any physical wave packet will generally have a nonzero overlap with the system's bound states, thereby violating this premise.
When the $S$ matrix exhibits poles in the first quadrant of the complex $k$ plane, time-growing bound states emerge.
Although an incident wave packet's initial overlap with these states may be negligible, this component undergoes exponential amplification, eventually dominating the system's long-time dynamics. 
Consequently, the wave-packet evolution deviates dramatically from the conventional scattering picture, rendering the transmission and reflection coefficients obtained from time-independent scattering methods unphysical.
 
Crucially, the breakdown of the scattering interpretation does not necessarily manifest through divergent reflection and transmission coefficients.
As demonstrated in Fig.~\ref{fig:01_time_independent_results}, these calculated quantities can remain finite even when the underlying scattering picture has failed.
Consequently, seemingly reasonable results from time-independent calculations can obscure the presence of time-growing bound states, potentially leading to fundamentally incorrect physical
   interpretations.
We therefore conclude that \textit{a priori} analysis of $S$-matrix pole distributions constitutes an essential prerequisite for applying time-independent scattering methods to non-Hermitian systems.

It is worth noting that the presence of time-growing bound states is an intrinsic characteristic of non-Hermitian scattering systems, independent of the specific profile or central wave number $k$ of the incident wave packet. 
As long as the system parameters are such that a pole appears in the first quadrant of the complex $k$ plane, any initial state with a nonzero overlap with the corresponding bound state will excite this growing mode, leading to a divergence in the system's response over time.

Furthermore, the presence of time-growing bound states suggests the necessity of incorporating nonlinear effects in certain parameter regimes.
Our analysis is based on the linear Schrödinger equation, which accurately describes the physics during the initial growth stage when the field intensity remains sufficiently small. 
However, as the field intensity grows exponentially within the system, nonlinear terms in the governing equations can no longer be neglected. 
These nonlinearities can play a crucial role in saturating the growth, ultimately leading to physically meaningful, stable solutions. 
Thus, a complete understanding of non-Hermitian systems in these regimes requires going beyond the linear approximation and incorporating nonlinear effects.

Finally, while our analysis focused on 1D tight-binding models, the underlying principles likely extend to a broader class of non-Hermitian systems, including those described by continuous Hamiltonians and higher-dimensional models. 
The key factor determining the applicability of time-independent scattering methods is the presence or absence of time-growing bound states, which can manifest in various non-Hermitian systems.

\begin{acknowledgments}
This work was supported by the Natural Science Foundation of the Jiangsu Higher Education Institutions of China, Grant No. 22KJB140009.
\end{acknowledgments}

\bibliography{MyLibrary.bib}

\end{document}